\documentstyle[12pt,aasms4]{article}

\begin{document}
\centerline{Submitted to the Astrophysical Journal}
\vskip 0.3in
\title{Evidence For Advective Flow From Multi-Wavelength Observations Of
Nova Muscae}

\author{\bf Ranjeev Misra}
\affil{Inter-University Centre for Astronomy and Astrophysics, Pune, India}
\authoremail{rmisra@iucaa.ernet.in}

\begin{abstract}
We model the UV/optical spectrum of the black hole binary Nova Muscae
as a sum of black body emissions from the outer region
of an accretion disk. We show for self-consistency that
scattering effects in this region are not important.
The black hole mass ($M \approx 6 M_\odot$), the inclination 
angle ($\mu \approx 0.5$) and the distance to the source 
($D \approx 5$ kpc) have been constrained by
optical observations during quiescence (Orosz et al. 1996). Using these values
we find that 
the accretion rate during
the peak  was ${\dot M} \approx  8 \times 10^{19}$ g sec$^{-1}$ and
subsequently decayed exponentially . 
 We define a radiative fraction
($f$) to be the ratio of the X-ray energy luminosity
to the total gravitational 
power dissipated for a keplerian accretion disk. We find that $f \approx 0.1$
and remains nearly constant during the Ultra-soft and Soft spectral states.
Thus for these states, the inner region of the accretion disk is 
advection dominated.
$f$ probably increased to $\approx 0.5$ during the Hard state and finally
decreased to $\approx 0.03$ as the source returned to quiescence.

We show that advective flow in the disk is optically thick due to high
accretion rates during the outburst. This is in contrast to some theoretical
models of advection dominated disks which require optical thinness.
We speculate that 
this optically thick advective disk could be the origin 
of the soft component if copious
external cold photons are available. The soft component could also
be due to a keplerian non-advective disk which terminates at around 
$R \approx 30$
Schwarzschild radius. However in this case the inner advective flow has
to be photon starved. 
Theoretical models of inner hot accretion disk are generally 
parameterized in terms of the
normalized accretion rate ${\dot m} = {\dot M}/{\dot M}_{Edd}$, where
${\dot M}_{Edd}$ is the Eddington accretion rate. Our results show that
Nova Muscae was in the Ultra-soft state when $ {\dot m} \ge 50$, in
the Soft state for $50  > {\dot m} > 2$ and in the Hard state
for $ {\dot m} \le 2$.

Our results constrain present and future theoretical models for
the inner regions of accretion disks around black holes. We 
highlight the need for multi-wavelength 
observations of future black hole novae to confirm the
results presented here. 
 
\end{abstract}


\section{Introduction}

The black hole X-ray binary Nova Muscae ( GS1124-683, GU Mucs )
was discovered 1991 January 9 by both the Ginga and 
Granat satellites ( Ebisawa et al. 1994; Gilfanov et al. 1993 and
references therein). The X-ray spectrum and evolution were similar
to other black hole X-ray Novae e.g. A0620-00 ( Elvis et al. 1975).
The X-ray luminosity had a rise time of $\approx 3 $ days and then
subsequently decreased exponentially with a decay time of $\approx 30$ days.
The source was observed to be in different spectral states during the
outburst (Ebisawa et al. 1994).
At the peak, the spectrum consisted of a soft component and a hard
X-ray power-law with photon spectral index of $\approx 2.5$. This 
state lasting till early February has been named the Ultra-soft 
state. The source was in the Soft state 
from mid-February to April where the hard X-ray
power-law nearly disappeared. There was a transition spectral
state during May, which was followed by the Hard state. In the
Hard state the spectrum consisted of a hard X-ray power-law with
photon spectral index of $\approx 1.5$ with the soft component
nearly absent. Details of these spectral states
and the temporal behavior of the source are given by Ebisawa et al. 1994.
X-ray observations by Sigma showed that the hard X-ray power-law
extended at least till $\approx 600$ keV whenever the source was detected
(Gilfanov et al. 1993). Nova Muscae was also observed by the 
{\it International Ultraviolet Explorer} (IUE) at several
epochs during the outburst and once by the {\it Hubble Space Telescope} (HST)
(Cheng et al. 1992). The energy spectral index in the UV band was observed to
be $\approx 0.3$ and the luminosity also decreased exponentially with
a decay time of $\approx 43$ days. Thus Nova Muscae was observed extensively
in both optical/UV and X-rays during the outburst.

Subsequent optical observation of the source during quiescence ( April 3
1992) revealed that the mass of the compact object is $\ge 3 M_\odot$
( Remillard, McClintock \& Bailyn 1992) confirming
that Nova Muscae is a black hole binary. Further extensive observations
during February 1993, refined the mass estimate. Ellipsoidal modeling
of the light curve indicate that the inclination angle  is in the range
$54^{o} < i < 64^{o}$ and that the black hole mass lies between
$ 5.0 M_\odot < M < 7.5 M_\odot$ (Orosz et al. 1996). Taking into
account the fraction of light from the accretion disk and the spectral
type of the secondary star, Orosz et al. 1996 estimate the distance
to the source to be $D = 5.5 \pm 1 $ kpc. With these constrains
and the extensive UV and X-ray observations, Nova Muscae is
the ideal source to test theoretical models. 
The spectral states of Nova Muscae are similar to those of other
black hole systems, including persistent sources like  Cygnus X-1.
Thus understanding this source will shed light on stellar black hole
systems in general.

Recently, two different models have been proposed
for Nova Muscae's Ultra-soft  and Soft spectral state by
Chakrabarti (1997) and Esin, McClintock, \& Narayan ( 1997). In
both these models the soft component arises from a non-advective
keplerian inner disk and the hard component arises from an advective flow.
The two models differ as to the location and nature of this 
advective flow. However, it is not clear if these X-ray models 
also predict the spectral shape and flux of the observed UV/optical
emission during these states.

In this paper, our motivation is to constrain these and other future
X-ray models by multi-wavelength analysis. We adopt a black hole
mass $ M = 6 M_\odot$, the inclination angle $\mu =0.5$ and 
distance to the source $D = 5$ kpc in accordance with the estimates
made by Orosz et al. (1996). The UV spectrum is fit as a sum of
black body emissions from the outer regions of an accretion disk. 
We show for self-consistency that scattering does not affect the
black body spectrum from this region. This fit fixes the accretion
rate at a given epoch. We then use the X-ray flux to infer the
radiative fraction $f$ defined to be the ratio of the observed radiative 
luminosity
to the luminosity expected from a non-advective keplerian accretion disk. 
We also place constrains on the location of any inner keplerian
disk which in Chakrabarti (1997) and Esin, McClintock, \& Narayan ( 1997)
models is the source of the soft component.

Cheng et al. (1992) have also modeled the UV/optical emission as black body
flux from the outer regions of an accretion disk.
They also constrain the disk parameters
using the X-ray flux. We
have extend their work with several modifications.
Unlike Cheng et al. (1992), we allow for the inner region of the
accretion disk to be advection dominated. We adopt specific values for
$M$, $\mu$ and $D$ instead of treating them like parameters.
We also constrain the model using
Ginga data ( 1 - 30 keV) rather than only 
Granat data which was from 3 keV onwards. 

In \S 1 we tabulate the specific observations used for this analysis.
In \S 2 the results are presented while in \S 4 we constrain and 
speculate on the source of the soft component. In \S 5 the paper is
summarized and the implications of the main results are discussed.

\section{The observations}

Nova Muscae was observed by the International Ultraviolet Explorer (IUE)
satellite at 12 epochs between January 17 and April 22, 1991 and
by HST on May 15, 1991 (Cheng et al. 1992). In this paper, we
consider four of the IUE and the HST observations covering the
different stages of the Nova decay (Table 1). We also consider
the CTIO observation on June 7 1991 (Bailyn 1992). Since the
spectral energy index for the HST and IUE observations were all
close to $\sim 0.3$, we assume that the spectral index had the same
value on June 7. 

\begin{table}
\caption{Observations of Nova Musca 1991}
\begin{tabular}{llll}
\hline
\hline
Date &  Optical/UV & X-ray \\
\hline
Jan 17 & IUE$^1$ & Ginga$^2$ \\
Feb 6  & IUE$^1$ & Ginga$^2$ \\
Mar 15 & IUE$^1$ & Ginga$^2$ (Mar 20)\\
Apr 22 & IUE$^1$ & Ginga$^2$ (Apr 19)\\
May 15 & HST$^1$ & Ginga$^2$ (May 17)\\
Jun 7 &  CTIO$^3$ & Ginga$^2$ (Jun 13)\\
\hline
\end{tabular}

$^1$ Cheng et al. (1992)
$^2$ Ebisawa et al. (1994)
$^3$ Bailyn (1992)

\end{table}

In the X-ray band, the source was frequently observed from January 8 1991
till October 17 1991 by Ginga ( Ebisawa et al. 1994). Six of these
observations were nearly simultaneous with the UV/optical data (Table 1)
and have been used in the analysis here. Ebisawa et al. (1994) fit
the X-ray data ( 1 - 30 keV) using a multi color disk model and a power law.
We have used this best fit spectrum as representative of the data.
Observation by Sigma showed that the hard X-ray power-law extended to
energies $ > 600$ keV ( Gilfanov et al. 1993). Thus we conservatively 
assume that the power-law extends to around 500 keV. We calculate
the luminosity of the source for $E > 1$ keV ($L_X$) using the
multi color disk model spectrum and the hard X-ray power-law extended
till 500 keV.

\section{ Results }

The optical/UV spectra of Nova Muscae has an energy spectral
index of 0.3. This indicates that this radiation is from
the outer region of a standard 
accretion disk with local black body emission ( Cheng et al. 1992). 
The local effective 
temperature of the disk is ( Shakura \& Sunyaev 1973),
\begin{equation}
T_s (r) \approx 7 \times 10^7 K\; ({M\over 6 M_\odot})^{-1/2} {\dot M}_{20}^{1/4}
 ({r\over r_s})^{-3/4} \chi^{1/4}
\end{equation}
where, ${\dot M}_{20}$ is the mass accretion rate in units of $10^{20}$ g 
s$^{-1}$, $r_s \equiv 2GM/c^2$ and $\chi = 1 - (3r_s/r)^{1/2}$. The
 flux at earth is,
\begin{equation}
F(E)  = {4 \pi E^3\mu  \over h^3 c^2 D^2} \int^{r_o}_{r_i} {1 \over 
exp(E/kT) - 1}\; r dr \;\;\hbox { ergs cm$^{-2}$ s$^{-1}$ ergs$^{-1}$}
\end{equation}
where $E$ is the energy of the photon, $\mu =$cos $i$,  $i$ being the
inclination angle and $D$ is the distance to the source. $r_o$ and
$r_i$ are the outer and inner radii of the accretion disk.
It should be noted
that $r_i$ is defined such that for $r < r_i$, Equation (1) is
no longer valid. In general the accretion disk may be even optically thick
and keplerian
for $r < r_i$  but scattering and Comptonization become important
in this region.  

The standard disk radiates as a sum of local black bodies if
scattering in the outer layers is negligible. This occurs if
the free-free frequency averaged absorption  opacity ($\kappa_{ff}$) is
larger or comparable to the scattering opacity ($\kappa_{es} = 0.4$ cm$^2$
g$^{-1}$). In the standard accretion disk theory (Shakura \& Sunyaev 1973),
\begin{equation}
{\kappa_{ff} \over \kappa_{es}} \approx 1.6 \times 10^{-4}
 {\dot M}_{20}^{-1/2} ({M \over 10 M_\odot})^{1/2} ({r\over r_s})^{3/4} > 1.0
\end{equation}
 Since $T_s(r_{UV}) \approx 3 \times 10^4$ K, the above constraint using
equation (1), translates to
\begin{equation}
{\kappa_{ff} \over \kappa_{es}} \approx 0.3 \;{\dot M}_{20}^{-1/4}\; ({T_s\over 
3 \times 10^4 \hbox {K}})^{-1} > 1.0
\end{equation}
Thus scattering does not strongly effect the UV continuum.

The outer regions of the accretion disk could be heated due
to X-ray irradiation. Vrtilek et al. 1990 have calculated the
spectrum from a disk where the irradiation dominates over
gravitational heating. They find that the spectrum is convex shaped
with the high energy spectral index being $-4/3$ and the low energy
one $\approx 1/3$. The spectral transition occurs at the energy  $E_T
\approx 3 k T(r_o)$ where$T (r_o)$ is the temperature of the outer
radius of the disk. The transition energy $E_T$ depends on the 
X-ray luminosity of the source. Further, the spectral index varies if
a stellar type spectrum (instead of black body) is used for the calculations.
The UV emission of Nova Muscae had a spectral index $\approx 0.3$ with
no breaks in the spectrum till $E \approx 10^{-7}$ ergs at least till
May. This will be consistent
with the irradiated disk model only if $T (r_o) > 6 \times 10^4 K$ all
through the observations. Detailed modeling of the irradiated disk
with local stellar-like spectrum is needed to confirm or rule out
this hypothesis. In this paper we have assumed that irradiation
effects are not important which is consistent with the observed
spectral shape.

The mass of the black hole, the inclination angle and the
distance to the source have been estimated by optical observations
during quiescence (Orosz et al. 1996).
In accordance with these results we have adopted $M = 6 M_\odot$,
$\mu =0.5$ and $D = 5$ kpc. Further we have assumed that the outer
radius
$r_o = 10^{11}$ cms. Since for all cases considered here,  
 $k T_s (r_o) << h\nu_{UV}$, the results are not sensitive to the
value of $r_o$ assumed. The optically thick disk emission (Equation 1)
also depends upon the accretion rate ($\dot M$) and $r_i$. If
$h\nu_{UV} << k T_s (r_i)$, then the UV spectrum determines $\dot M$
and the  X-ray observations constrain $r_i$. We
show the results of such a fitting in Figure 1. For January 17 the
accretion rate turns out to be $8 \times 10^{19}$
g s$^{-1}$ in order to match the UV emission. 
The minimum value of $r_i$, which is not in conflict
with the X-ray observations (at $E = 1$ keV) is $r_{min} = 63 r_s$. 
In general, 
$r_i$ should be larger than $r_{min}$ and the spectrum should flatten
somewhere in the energy range $.01$ keV $ < E < 1$ keV. Thus, setting
$r_i = r_{min}$ represents the maximum power the system can generate
for $E < 1$ keV. We find that all through the outburst $r_{min}$ varied
from $30$ to $65 r_s$. 

The total gravitational energy per second that should be 
released for a keplerian disk in the region $r < r_i$ is,
\begin{equation}
L_G ( r_i ) = L_{GM}\; [1 - {9 r_s\over r_i}(1 - {2\over 
\sqrt{3 r_i/r_s}})] 
\end{equation}
where $L_{GM} = {\dot M} c^2/12$ is the total gravitational energy
released. A fraction of this energy is radiated while the rest
is advected into the black hole. We define a radiative fraction 
\begin{equation}
f = {L_X\over L_G (r_i)}
\end{equation}
where $L_X$ is the observed X-ray ($E > 1$ keV) luminosity of the source.
$f << 1$ would indicate that the inner regions of the disk
is highly advective. Since in general $r_i > r_{min}$, the maximum value of this
fraction is,
$f_{max} = {L_X/ L_G (r_{min})}$. The minimum value is obtained
by setting $r_i >> r_s$ i.e.  $f_{min} = {L_X/ L_{GM}}$. In
Figure 2 we show these limits on $f$ for different times of the outburst.
Note that $f$ is well constrained by these limits and that the result
is compatible with $f \approx 0.1$ for the six epochs. These epochs
correspond to the Ultra-soft, Soft and the Soft
to Hard transition states. Figure 3 shows the variation
of the extremal values of $f$ with distance. Solid lines
are for $M = 6 M_\odot$, while dotted line is for $M = 10 M_\odot$.

Figure 4 shows the inferred accretion rate from the UV data. As
noted by Cheng et al. (1992), $\dot M$ varies exponentially with
time. We define a normalized rate ${\dot m} \equiv {\dot M}/{\dot M}_{Edd}$
where ${\dot M}_{Edd} = 4\pi GMm_p/c\sigma_T \approx 
 8.5 \times 10^{17}$ g sec$^{-1}$ is the Eddington accretion rate. 
Nova Muscae was in the Ultra-soft state when $ {\dot m} \ge 50$, in
the Soft state for $50  > {\dot m} > 2$ and in the Hard state
for $ {\dot m} \le 2$. At the peak of the outburst ${\dot m} \approx 100$.

We extrapolate this variation of ${\dot M}$ to the Hard state where
UV data is not available. We then repeat the analysis 
using all the X-ray observations made by Ginga ( Ebisawa et al. 1994)
i.e. not only the ones listed in Table 1. Figure
5 shows the radiative fraction which has been 
calculated assuming that the accretion rate decreases
exponentially. During the Hard state the
fraction is closer to unity while it becomes much less than
unity again as the source returns to quiescence. 
This result depends on the assumption that $\dot M$ decreased exponentially
all through the decay.

\section{The Soft Component}

Since the radiative fraction is close to 10 \% for the first
100 days of the outburst, the flow in the inner region is
advection dominated. Such advection dominated flows are
characterized by nearly free-fall speeds i.e. the radial
velocity $v_r \le c (r_s/r)^{1/2}$. In steady state, 
${\dot M} = 4\pi r h \rho v_r$, where $\rho$ and $h$ are
the mass density and local half-height of the disk respectively.
Since most of the matter is flowing advectively, this lead to
\begin{equation}
\tau \ge 26.5\; {\dot M}_{20} \;({r\over 5 r_s})^{-1/2}
\end{equation}
where $\tau$ is the scattering optical depth in the vertical direction. 
Thus during the Ultra-soft and Soft states, the 
inner region of the disk is optically thick. This is surprising
since advective flow have generally been modeled to be optically
thin and hard X-ray producing regions. The Compton y parameter
for such a region is,
\begin{equation}
y \ge 5.5\; {\dot M}^2_{20}\; ({r\over 5 r_s})^{-1}\; ({T\over 10^7 K})
\end{equation}
At least for the Ultra-soft state, $y > 1$ which would lead to saturated
Comptonization of any external cold photons. The emergent
spectrum should be a Wien peak. It is attractive to identify
this emission with the soft component observed in the source. 
However, it is not clear whether the soft component can be fitted
to a single temperature Wien peak since the data is well
described by a multi-color disk emission (Ebisawa et al. 1994).
Perhaps the inner region has temperature variations leading
to an emergent spectrum which is a sum of local Wien peaks
which might resemble a multi-color disk emission.
Moreover, one needs to specify the source of copious soft 
photons to be Comptonized. Further, the production of hard X-rays
has to be incorporated into this model.

The soft component has generally been interpreted as a 
multi-color disk emission from a keplerian optically thick disk
(Ebisawa et al. 1994, Misra \& Melia 1997). Ebisawa et al. (1994)
fit the data to such a model which has two parameters namely,
the color temperature ($T_c$) of the innermost radius, and the 
total luminosity 
of the component ($L_s$). Assuming that $T_c = T_{eff}$, where $T_{eff}$ is
the effective temperature, an empirical inner radius $r_e$ can be calculated
from $T_c$ and $L_s$. During the Ultra-soft state $r_e \approx 7 \times 10^6$ cms for $ D = 5$ kpc and $\mu =0.5$ (Ebisawa et al. 1997). 
However, since $T_c > T_{eff}$ due to Compton
scatterings the physical inner radius $r_{is} > r_e$. Defining the color
factor $f_c \equiv T_c/T_{eff}$, $r_{is} \approx r_e f_c^2$. 

We first consider the case where $r_{is} \approx 3 r_s = 5.4 
\times 10^6$ cms i.e. an
keplerian disk which extends till the last stable orbit. This
corresponds to the unlikely value of $f_c \approx 1.0$. 
Now $L_s \approx
G M{\dot M_s}/2 r_{is}$, where ${\dot M_s}$ is the accretion rate
in the optically thick disk. Since $L_s \approx 6 \times 10^{38}$ ergs 
sec$^{-1}$ during the peak of the outburst, the maximum  ${\dot M}_s \approx 
8 \times 10^{18}$ g s$^{-1}$. However, during the peak of the outburst
the total accretion rate was ${\dot M} = 8 \times 10^{19}$ g s$^{-1}$. Thus
only 10\% of the accreting matter is in the
form of a keplerian disk. The rest of the accretion must
be flowing advectively on top. As mentioned above
this overlying flow
should be optically thick and should alter the nearly black body emission
of the keplerian disk. Such a modification should
necessarily lead to $f_c > 1$. Thus if a keplerian disk is the source
of the soft component it is unlikely that it extends till the last stable
orbit. Thus models with keplerian disks with overlying corona (Liang
\& Price 1977, Haardt \& Maraschi 1993) do not seem to be applicable
to the Ultra-soft state of Nova Muscae.

We now consider the situation when $r_{is} > 3 r_s$. If there is
no overlying advective flow, $r_{is} \approx 5.4 \times 10^7 \approx 30 r_s$,
in order that the power released matches the observed $L_s$. In this case, $f_c
\approx 3$. It should be noted that
theoretically, $f_c \approx 1.8$ and is nearly independent of disk 
parameters (Shimura \& Takahara 1992). It is not clear why $f_c$ should
be greater than this theoretical prediction. The inner accretion 
disk $ r < 30 r_s$ has to be advection dominated with radiative fraction
$f << 0.1$. 
Models with an outer keplerian disk with the inner region being advection
dominated have been invoked to explain the spectrum of black hole
candidates (for e.g. Shapiro, Lightman \& Eardley 1976, 
 Misra \& Melia 1996, Chakrabarti 1997,
Esin, McClintock \& Narayan 1997). These
models differ in the radiative mechanism producing the hard X-rays,
geometry and the presence of shocks in the advective flows. However,
the accretion rate used in these models are lower than the one inferred
in this paper. For example, Chakrabarti 1997  calculate the spectrum
for ${\dot m} \equiv {\dot M}/{\dot M}_{edd} \approx 1$, 
while the results obtained here imply that
${\dot m} \approx 100$. Thus, it is not clear if
the results of these papers will be valid for such high
accretion rates. In particular, as mentioned above ${\dot m} \approx 100$
implies that the advection dominated disk should be optically thick and
not optically thin as has been assumed for these models. The transition
from an outer keplerian disk to an inner hot region could
be occurring at a radius when the disk is radiation pressure dominated 
and locally unstable (Shapiro, Lightman \& Eardley 1976). This occurs
at,
\begin{equation}
r_{tr} \approx 350 r_s \alpha^{2/21}\; ({M\over 6 M_\odot})^{-2/3} {\dot M}_{19}^{16/21}
\end{equation}
This is consistent with the results presented here as $r_{min} \approx 50 r_s
< r_{tr}$ all through the outburst.

Disks which are advection dominated but optically thick
have been studied by Abramowicz et al. (1988) and are called
slim accretion disks. These disks have low viscosity and
high accretion rates ${\dot m} >> 1$. Note that $\dot m$
defined by Abramowicz et al. (1988) is a factor of 16 times
greater than the one used here. Abramowicz et al. (1988)
calculate that the radiative fraction $f \approx 0.1$ only
for ${\dot m} > 500$ for $\alpha = 0.001$, where $\alpha$
is the viscosity parameter. Since the normalized accretion rate
calculated in this paper, ${\dot m} \approx 100$ the 
slim disk model has to be modified to be compatible with these
results.

\section{Summary and Discussion}

Optical observations of Nova Muscae during the quiescent state, have
constrained the mass of the black hole $M \approx 6 M_\odot$, the
inclination angle $\mu \approx 0.5$ and distance to the source $D \approx 
5$ kpc (Orosz et al. 1996). Using these parameters we find that:

\noindent 1) The optical/UV spectrum during the outburst can be
self-consistently modeled by sum of black body emission from the
outer regions of an accretion disk. The accretion rate during
the peak of the outburst, ${\dot M} \approx 8 \times 10^{19}$
g sec$^{-1}$. The accretion rate then decreased exponentially with
time at least till the source was making a transition from the
Soft to the Hard state. 
The source was in the Ultra-soft state when $ {\dot m} \equiv 
{\dot M}/{\dot M}_{edd} \ge 50$, in
the Soft state for $50  > {\dot m} > 2$ and in the Hard state
for $ {\dot m} \le 2$.

\noindent 2) We define a radiative fraction $f$ to be 
the ratio of the
total radiative power emitted to the total expected gravitational power
for a standard accretion disk. We find that $f \approx 0.1$ during
the Ultra-soft, Soft and Soft to Hard transition states. The presence
of an advective component is implied. If the
accretion rate decreased exponentially all through the outburst,
$f$ increased to $ \approx 0.5$ during the Hard state before finally
decreasing to $\approx 0.03$ as the source returned to the quiescent state.

\noindent 3) During the peak of the outburst (when the source
was in the Ultra-soft state) the high accretion rate 
(${\dot M} \approx 8 \times 10^{19}$ g sec$^{-1}$) implies that advective
component of the disk has to be optically thick to scattering.
The soft component may then have it's origin in this optically
thick but advective zone if there are sufficient external photons
to be Comptonized. 

\noindent 4) A keplerian disk if present does not extend to
the last stable orbit during the Ultra-soft state. The X-ray soft component
emission during the Ultra-soft state could be due to a
keplerian disk terminated at a radius $r \approx 30 r_s$, with
a highly advective photon-starved flow in the inner region.

We point out that the accretion rate during outburst is
${\dot m} \approx 100$. The recent
models for Nova Muscae (Esin, McClintock \& Narayan
1997; Chakrabarti 1997)  do not consider such high accretion
rates and hence their results cannot be directly compared to those
obtained here. Nevertheless the basic picture of these
models i.e. an outer keplerian disk with an inner
highly advective disk is consistent with the results.

It should be noted that a low radiative fraction does not necessarily
mean that the radiative mechanism for the inner disk in inefficient.
It could also be possible that the disk dissipates less than a Keplerian
one, and most of the gravitational energy is converted to radial
motion instead of thermal.

In this paper, we have assumed that the X-ray irradiation effects
on the disk are not important. The constancy of the UV spectral index
and the absence of a spectral break indicates that this assumption is
probably valid. However, more detailed modeling of irradiated disks
and broad band spectra are needed before any conclusive statements can be
made.
The results presented in this paper are based on the observations
of the source during the quiescent state and the identification of
the UV/optical spectrum (during the outburst) as a 
sum of black body emission from the outer
regions of an accretion disk. Theoretical models for X-ray Nova which
do not predict such high accretion rates would need to self-consistently
incorporate an alternate
explanation for the UV spectrum.  
We underline the need for more sensitive multi-wavelength
observations of future X-ray Nova to confirm the results presented here.

\acknowledgments
The author thanks F. Melia for useful discussions and comments.

\clearpage
 
\figcaption{The multi-wavelength observations of Nova Muscae
for January 17. The UV spectrum is fit by a sum of
black body emission from the outer region of the accretion disk.
The inner radius of such a black body emitting disk is chosen
such that the spectrum is not in conflict with X-ray observations.
\label{fig1}} 

\figcaption{The upper and lower limits of the 
radiative fraction $f$ for the six simultaneous
multi-wavelength observations described in Table 1. \label{fig2}}

\figcaption{The variation of the radiative fraction $f$
with distance. $\mu = 0.5$ is fixed. Solid line: $M = 6 M_\odot$; Dotted line:
$M = 10 M_\odot$. For the rest of the paper,  $M = 6 M_\odot$
and $D = 5$ kpc  has been adopted.\label{fig3}}

\figcaption{ Variation of the accretion rate ${\dot M}$ as a function
time for the six set of simultaneous observations (points). The solid
line is the best fit to these points.  \label{fig4}}

\figcaption{ The upper and lower limits of the 
radiative fraction $f$ using the best fit line of Figure 4 and extending
it to later dates.
\label{fig5}}

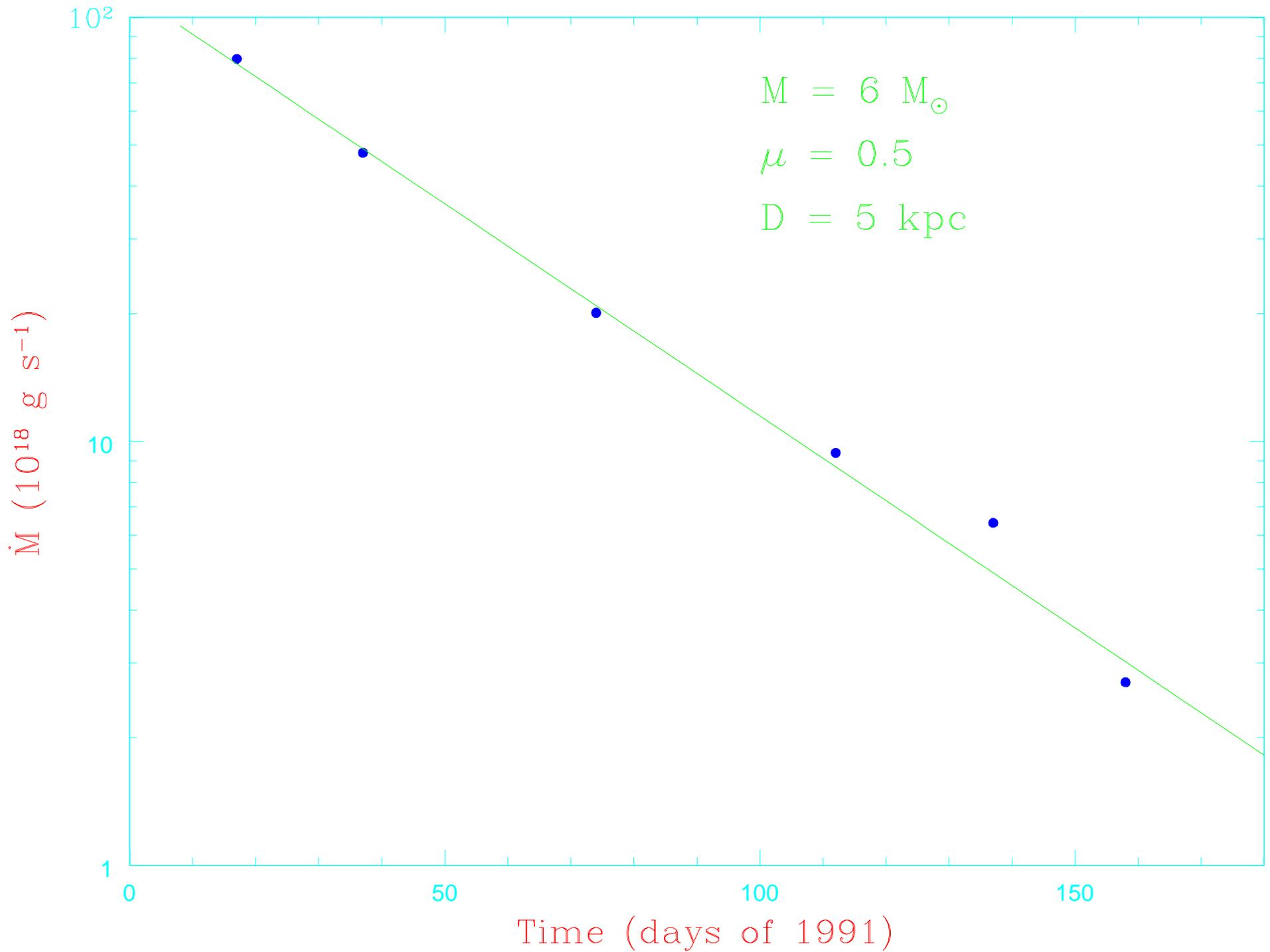
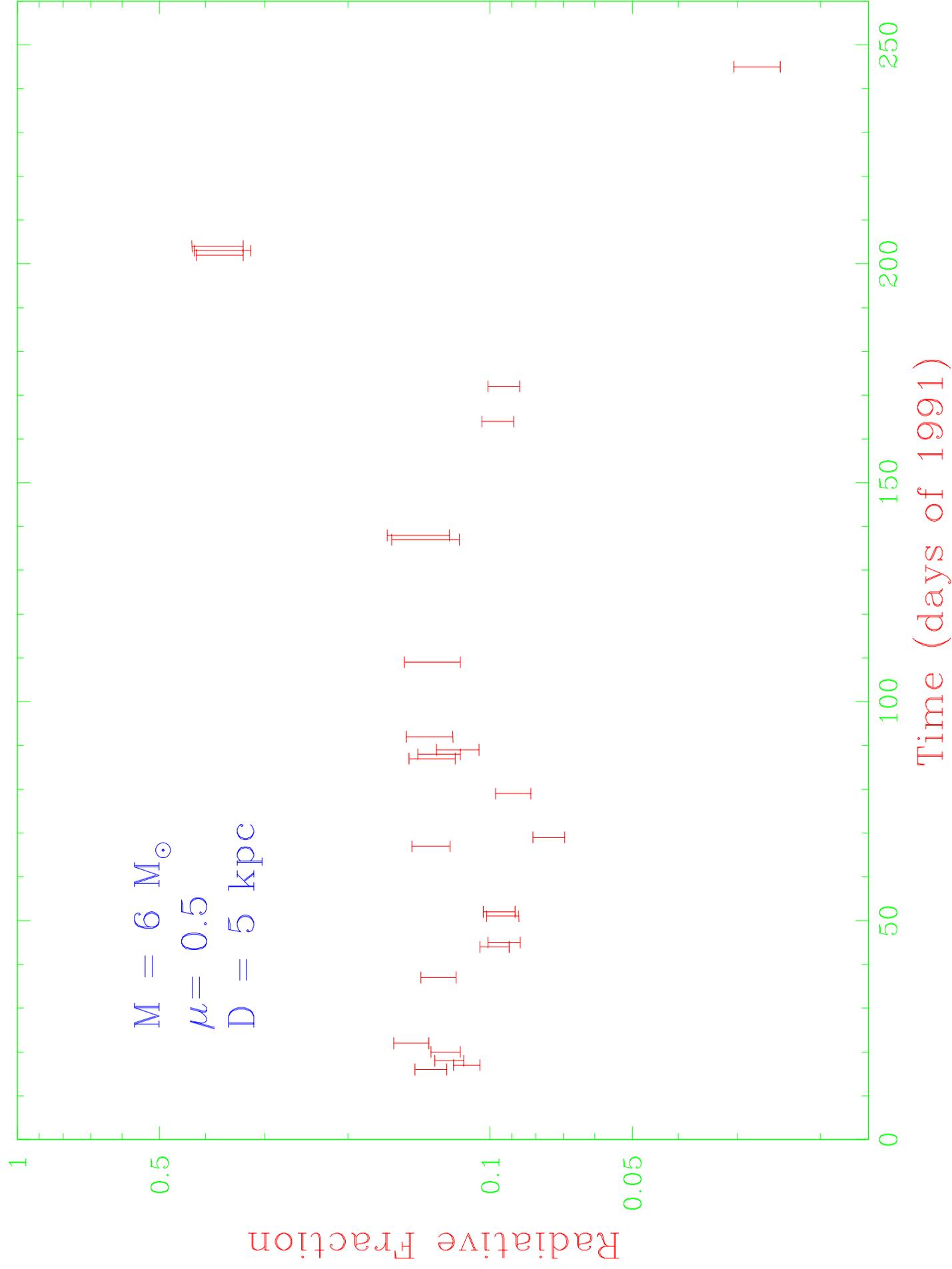
\end{document}